\documentclass[12pt]{article}
\begin{document}
\title{The Modified Dirac Equation}
\author{B.G. Sidharth\\
International Institute for Applicable Mathematics \& Information Sciences\\
Hyderabad (India) \& Udine (Italy)\\
B.M. Birla Science Centre, Adarsh Nagar, Hyderabad - 500 063
(India)}
\date{}
\maketitle
\begin{abstract}
We consider the behavior of the particles at ultra relativistic
energies, for both the Klein-Gordon and Dirac equations. We observe
that the usual description is valid for energies such that we are
outside the particle's Compton wavelength. For higher energies
however, both the Klein-Gordon and Dirac equations get modified and
this leads to some new effects for the particles, including the
appearance of anti particles with a slightly different energy.
\end{abstract}
\section{Introduction}
It is now being recognized that at very high energies, due to the
effects of non commutative geometry, the usual energy momentum
dispersion relation
\begin{equation}
E^2 = p^2 + m^2\label{e1}
\end{equation}
gets modified (we use natural units, $c = 1 = \hbar$). This should
result in the modification of the Klein-Gordon and Dirac equations
(Cf.ref.\cite{bgsijtp,bgsijmpe,tduniv} and several references
therein). We would like to study this aspect in greater detail.
Before we do so however, let us reexamine some well known aspects of
the Klein-Gordon and Dirac equations.\\
Our starting point is the well known Klein-Gordon equation
\cite{fresh,schweber}
\begin{equation}
\sum_\mu [D^2_\mu - k^2] \psi = 0,\label{2.6}
\end{equation}
where
\begin{equation}
D_\mu = \partial / \partial x_\mu - (\imath e / \hbar c) A_\mu \quad
\mbox{and} \quad k = mc/\hbar .\label{2.7}
\end{equation}
Equations (\ref{2.6}) and (\ref{2.7}) are obtained from (\ref{e1})
with the usual substitutions and the introduction of the
electromagnetic field.\\
The difficulties with the Klein-Gordon (K-G) equations are well
known. They arise from the fact that equation (\ref{2.6}) is second
order in the time derivative, unlike the Schrodinger equation. These
difficulties however, as was later shown could be circumvented if
the K-G equation were to be interpreted in a field theory context.\\
However valuable insight about (\ref{2.6}) can be obtained if it can
be recast in the Hamiltonian form.\\
We follow Feshbach \cite{fresh} for this purpose and look for an
equation like,
\begin{equation}
H \Psi = \imath \hbar (\partial \Psi / \partial t).\label{2.10}
\end{equation}
We then make the substitution
\begin{equation}
\psi_4 = - k^{-1} D_4 \psi\label{2.11}
\end{equation}
to get
$$D_4 \psi + k \psi_4 = 0,$$
\begin{equation}
\sum_k D^2_k \psi - kD_4\psi_4 - k^2\psi = 0.\label{2.12}
\end{equation}
We further make the substitutions
$$\psi = 1 / \sqrt{2} (\psi + \chi ),$$
\begin{equation}
\psi_4 = 1 / \sqrt{2} (\psi - \chi).\label{2.13}
\end{equation}
Thus in effect we have considered the K-G wave function $\Psi$ to be
a two component object,
\begin{equation}
\Psi = \left(\begin{array}{ll} \psi \\ \chi
\end{array}\right),\label{2.16}
\end{equation}
We then get
$$\imath \hbar (\partial \psi / \partial t) = (1/2m) (\hbar / \imath
\nabla - eA/c)^2 (\psi + \chi)$$
$$\quad \quad \quad \quad + (e \phi + mc^2 \chi$$
$$\imath \hbar (\partial \chi / \partial t) = - (1/2m) (\hbar / \imath
\nabla - eA/c)^2 (\psi + \chi)$$
\begin{equation}
\quad \quad \quad \quad + (e \phi - mc^2 \psi.\label{2.15}
\end{equation}
It is interesting to note that in (\ref{2.15}) we have the following
symmetry:
$$t \to -t, \, \chi \to \phi , \, \phi \to \chi , \, A_4 \to - A_4$$
Because of the description (\ref{2.16}), we now have to introduce
the Pauli matrices for defining a suitable current and continuity
equations.\\
Also the normalization of $\Psi$ is now given by
\begin{equation}
\int \Psi^* \sigma_3 \Psi d^3 x = \pm 1,\label{2.24a}
\end{equation}
As Feshbach and Villars note "... the increase of the 'degrees of
freedom' connected with the appearance of a second-order time
derivative in the Klein-Gordon equation corresponds to the
simultaneous description of a particle of either positive or
negative charge; i.e., the value of the charge becomes a degree of
freedom of the system. The solution describing a particle of
positive charge may be normalized to $+ 1$; the charge conjugate
solution will automatically be normalized to $(-1)$ and thus
describe a negative charge."\\
However in the absence of an external electromagnetic field,
(\ref{2.15}) splits into two separate equations, one describing the
positive solution and the other describing the negative solution.\\
To get better insight into this circumstance let us write
$$\Psi = \left(\begin{array}{ll} \psi_0 (p) \\ \chi_0 (p)
\end{array}\right) e^{\imath / \hbar (p \cdot x - Et)}$$
\begin{equation}
\Psi = \Psi_0 (p) e^{\imath / \hbar (p \cdot x - Et)}\label{2.25}
\end{equation}
We consider separately the positive and negative values of $E$
(coming from (\ref{e1}), viz.,
\begin{equation}
E = \pm E_p ; \quad E_p = [(cp)^2 + (mc^2)^2]^{\frac{1}{2}}
.\label{2.26}
\end{equation}
The solutions associated with these two values of $E$ are
$$\phi_0^{(+)} = \frac{E_p + mc^2}{2(mc^2E_p)^{\frac{1}{2}}}$$
$$\chi_0^{(+)} = \frac{mc^2 - E_p}{2(mc^2E_p)^{\frac{1}{2}}},$$
for $E = E_p$ and
$$\phi_0^{(-)} = \frac{mc^2 - E_p}{2(mc^2E_p)^{\frac{1}{2}}}$$
$$\chi_0^{(-)} = \frac{E_p + mc^2}{2(mc^2E_p)^{\frac{1}{2}}} \, ,$$
for $E = -E_p$.\\
As is well known the positive solution $(E = E_p)$ and the negative
solution $(E = -E_p)$ represent solutions of opposite charge. It is
also well known that in the non relativistic limit the $\chi$
components are reduced with respect to the $\phi$ components, by the
factor $(p / mc)^2$. We also mention the well known fact that a
meaningful subluminal velocity operator can be obtained only from
the wave packets formed by positive energy solutions. However the
positive energy solutions alone do not form a complete set, unlike
in the non relativistic theory. This also means that a point
description in terms of the positive energy solutions alone is not
possible for the K-G equation, that is for the position operator,
$$\delta \left(\vec{X} - \vec{X}_0\right)$$
In fact the eigen states of this position operator include both
positive and negative solutions. All this is well known
(Cf.ref.\cite{fresh,schweber,bd}).\\
The point is that if we approach distances of the order of the
Compton wavelength, the negative energy solutions begin to dominate,
and we encounter the well known phenomenon of Zitterbewegung. This
modifies the coupling of the positive solutions with an external
field, particularly if the field varies rapidly over the Compton
wavelength. In fact this is the origin of the well known Darwin term
in the Dirac theory \cite{bd}. The Darwin term is a correction to
the interaction of the order
\begin{equation}
\left(\frac{p}{mc}\right)^4 \quad \mbox{and} \, \,
\left(\frac{p}{mc}\right)^2\label{X}
\end{equation}
for spin $0$ and spin $1/2$ particles respectively.
\section{Modified Equations}
We will now expand our outlook and consider the effects of a non
communicative geometry or equivalently a fuzzy space time, all of
which is symptomatic of the fact that we cannot go down to point
space time. Then as has been argued in detail \cite{tduniv} the
dispersion relation (\ref{e1}) gets modified and becomes the
Snyder-Sidharth Hamiltonian
\begin{equation}
E^2 = p^2 + m^2 + \alpha l^2 p^4 \quad \left(c = 1 = \hbar
;\right)\label{14}
\end{equation}
We can see that $\alpha \sim 0 (1)$. This follows by noting that
(\ref{14}) gives an extra term giving effects of Zitterbewegung at
the Compton wavelength. Thus it is like the Darwin term $\sim l^2
p^4$ encountered in (\ref{X}) \cite{fresh}. Hence $\alpha \sim 0
(1)$.\\
We next observe that effectively
$$m^2 \to \vec{m}^2 = m^2 + \, \alpha \, l^2 p^4$$
So we can get as in the usual theory of the Dirac equation
\begin{equation}
\left(\gamma^0 p^0 + \Gamma + \bar{m}\right) \psi = 0\label{15}
\end{equation}
where  $$\bar{m} \approx m + \frac{1}{2} \alpha l^2 p^4$$ We can
demonstrate this as follows. We first rewrite (\ref{15}) as
\begin{equation}
\left(\frac{\partial}{\partial t} - \sum^{3}_{k = 1} \alpha^k
\frac{\partial}{\partial x^k} - \imath m \beta - \lambda l p^2
\right) \psi = 0\label{16}
\end{equation}
Let us now left multiplied by
$$\left(\frac{\partial}{\partial t} + \sum \alpha^k
\frac{\partial}{\partial x^k} + \imath m \beta + \lambda l p^2
\right)$$ to get
$$\left(D - \lambda^2 l^2 p^4 \right) \psi = 0$$
where $D$ is the full $K-G$ Hamiltonian and where we choose
$$\alpha^k \lambda + \lambda \alpha^k = 0, \quad \beta \lambda +
\lambda \beta = 0$$ Going over to the usual $\gamma$ matrices,
$$\gamma^0 = \beta , \quad \gamma^k = \beta \alpha^k = \gamma^0
\alpha^k ,$$ we get, with $$\Theta = \gamma^0 \lambda , \quad \Theta
= \gamma^5$$ whence the so called Dirac-Sidharth equation (\ref{16}).\\
There is a correction for the mass in (\ref{16}), but which is a
non-invariant
under reflection.\\
This indicates a decay of the Fermion at ultra high energy -- it is
akin to "splitting" of mass a la the Zeeman effect, as will be
discussed
below.\\
If $m = 0$ to start with, (\ref{16}) shows the origin of the
neutrino mass. Earlier Chados and others inserted this term ad hoc
and interpreted it as a superluminal neutrino. These ideas were
subsequently dropped \cite{chen}.\\
Let us analyse the ultra high energy Dirac equation (\ref{16})
\cite{bgsijmpe} further. This can be done best, as for the
Klein-Gordon case above or for the usual Dirac equation by
considering a reduction to the Pauli equation as is usually done
\cite{bd}.\\
We get, with
$$\psi = e^{\frac{\imath}{\hbar} mc^2 t} \left(\begin{array}{ll} \phi \\ \chi \end{array}
\right),$$ for the modified Dirac equation, this time
$$\phi = - \frac{\{c(\sigma \cdot \pi) + \omega \}}{2mc^2 + eV}
\chi$$ with the Pauli-like equation for $\chi$ (instead of $\phi$ as
in the usual theory), given by
\begin{equation}
\imath \hbar \dot{\chi} = [eV - \frac{\{c (\sigma \cdot \pi)\}^2 -
\omega^2}{2mc^2 + eV}] \chi\label{17}
\end{equation}
where $\omega = \Theta lp^2$. For a magnetic field $\vec{B}$, this
throws up the spin - magnetic field coupling, this time given by,
remembering that we have
$$c(\sigma \cdot \pi)^2 = c^2 \pi^2 - \frac{e\hbar c^2}{c}
\vec{\sigma} \cdot \vec{B}$$
$$ec\hbar (\vec{\sigma} \cdot \vec{B} + \frac{\omega^2}{ec\hbar})$$
instead of $ec\hbar (\vec{\sigma} \cdot \vec{B})$ for the usual
Dirac equation. We can see from (\ref{17}) that we now have,
additionally a spin half particle with charge $-e$, but the spin
magnetic energy shifted due to the additional term involving
$\omega$. The other, positive energy solution $\phi$ of the usual
theory, represents a spin half particle of charge $e$, but a
different spin magnetic coupling energy with the opposite sign of
$\omega^2$. In any case this too differs from the usual Dirac theory
due to the new effect of the $\omega$ term, that is due to the
Hamiltonian (\ref{14}).\\
The "splitting" can be seen because of the new $\omega$ term in
(\ref{16}). If
$\omega$ were $0$ we would be back to the usual theory.\\
We have here, as noted elsewhere  a situation rather like that
corresponding \cite{bgsijmpe} to the (neutral) $k^0$ meson in the
following sense. The positive and negative solutions are given as in
the K-G case, as is well known by \cite{bd}
$$\psi_1^+ = \left[ \begin{array}{ll} 1 \\ 0 \\ 0 \\ 0\end{array}\right] e^{+\frac{\imath}{\hbar} (p \cdot x -
Et)} \quad  \psi_2^+ = \left[\begin{array}{ll} 0 \\ 1 \\ 0\\
0\end{array} \right] e^{+\frac{\imath}{\hbar} (p \cdot x - Et)}$$
$$\psi_1^{(-)} = \left[ \begin{array}{ll} 0 \\ 0 \\ 1 \\ 0\end{array}\right] e^{-\frac{\imath}{\hbar} (p \cdot x -
Et)} \quad  \psi_2^{(-)} = \left[\begin{array}{ll} 0 \\ 0 \\ 0 \\
1\end{array} \right] e^{-\frac{\imath}{\hbar} (p \cdot x - Et)}$$ If
we consider amplitudes like $<\psi_1^{(+)} \psi_1^{(-)} >$, they are
not real but rather contain the imaginary (Zitterbewegung) terms as
can be easily verified, and these terms again vanish on taking
averages over time intervals of the order of the Compton time
\cite{diracpqm,cu}. That is, for energies high enough to penetrate
the Compton scale, as we are considering here, we have complex
amplitudes, indicative of non conservation of probability and
decay.\\
As can be seen, in this ultra relativistic case in which we are at
the Compton wavelength and therefore encounter subsequently the
negative energy solutions, there is a new negative energy particle,
with a slight shift of energy. This is exactly as for the K-G
equation, as noted in Section 1: there is an increase in the number
of degrees of freedom with a description for positive charge and
another for negative charge. The situation is reminiscent of the
spectral splitting in the Zeeman effect \cite{powell}. At the
relativistic energies, we have the usual Dirac equation. This
equation is meaningful only for minimum distances greater than the
Compton wavelength. At ultra high energies there is a "splitting"
due to the term $lp^2 \gamma^5$. In other words this new effect is
characterized by the appearance of an "anti" particle.
\section{Remarks}
1. We stress again that the description in terms of positive energy
solutions alone, as in conventional theory is possible only outside
the Compton wavelength of the particle. At higher energies, as seen
above, we encounter negative energy solutions and a new effect
namely an anti particle with a slightly different energy as compared
to the original particle. To put it another way, the extra term in
the SS Hamiltonian (\ref{14}) is analogous to the magnetic field
which leads to the Zeeman "splitting". It is interesting that in
this decay the $\gamma^5$ matrix of the weak interaction plays a
role. In this context, we should note once again that the positive
energy solutions alone do not form a complete set and the connection
between a completeness,
unitarity and conservation of probability is all too well known \cite{Gasiorencz}.\\
2. If we had started with a mass zero particle, then the SS
Hamiltonian provides a mass. Indeed it has been explicitly shown
\cite{glinka} that the Hamiltonian can be approximated by
$$\frac{p^2}{2m} + m ,$$
which shows the mass term explicitly.\\
In this connection it may be observed that as noted above, Chodos
\cite{chodos} had ad hoc proposed an equation similar to the
massless modified Dirac equation. This started off much work on so
called superluminal neutrinos, but finally the consensus seems to be
that such neutrinos
do not exist \cite{yangijmpa}.\\
It may be argued that the mass generating term seen above endows
Goldstein bosons with a mass. Indeed the author has argued for many
years for a photon has a mass $\sim 10^{-65}gms$, well within the
experimental upper bound \cite{bgs1,bgs2,bgs3}. This can be argued
from the SS Hamiltonian itself. In any case Rao \cite{rao} had
experimentally concluded that the photon has a particle, rather than
a wave nature. It is proposed to repeat these experiments to verify
all
this.\\
4. It is interesting to note the following symmetry for the modified
Dirac equation:
$$t \to -t \to \phi \to \chi$$
as indeed for the Klein-Gordon equation seen earlier.\\
5. Finally it may be observed that the two two-component solutions
of the Dirac equation seen above, viz., $\phi$ and $\chi$ could be
thought of as in an isospin formalism. Thus the four component Dirac
wave function consisting of $\phi$ and $\chi$ represents a single
particle for the usual energies, much as the proton and neutron
could be thought of as the same particle in the absence of the
electromagnetic interactions. It is only when the extra term in the
SS Hamiltonian is introduced that the up and down states are
distinguishable.

\end{document}